\documentclass{INTERSPEECH2023}
\usepackage{multirow,amssymb,marvosym}

\interspeechcameraready

\title{Exploring the Interactions between Target Positive and Negative Information for Acoustic Echo Cancellation}
\name{Chang Han$^{1,\dagger}$, Xinmeng Xu$^{1,\dagger}$
, Weiping Tu$^{1,2,3}$\textsuperscript{,\Letter}
, Yuhong Yang$^{1,3}$, Yajie Liu$^1$ \thanks{$\dagger$ Equal contribution} \thanks{\Letter 
 Corresponding author} 
\thanks{This work was supported in part by the National Nature Science Foundation of China (No. 62071342, No.62171326), the Special Fund of Hubei Luojia Laboratory (No. 220100019), the Hubei Province Technological Innovation Major Project (No. 2021BAA034) and the Fundamental Research Funds for the Central Universities (No.2042023kf1033). The numerical calculations in this paper have been done on the supercomputing system in the Supercomputing Center of Wuhan University.}} 
\address{
  $^1$NERCMS,
School of Computer Science, Wuhan University, China\\
  $^2$Hubei Luojia Laboratory,  China\\
  $^3$Hubei Key Laboratory of Multimedia and Network Communication Engineering, \\Wuhan University, China
  }
\email{\{changhan, xuxinmeng, tuweiping, yangyuhong, yajieliu07\} @whu.edu.cn}

\begin{document}

\maketitle
 
\begin{abstract}
Acoustic echo cancellation (AEC) aims to remove interference signals while leaving near-end speech least distorted. As the indistinguishable patterns between near-end speech and interference signals, near-end speech can't be separated completely, causing speech distortion and interference signals residual. We observe that besides target positive information, e.g., ground-truth speech and features, the target negative information, such as interference signals and features, helps make pattern of target speech and interference signals more discriminative. Therefore, we present a novel AEC model encoder-decoder architecture with the guidance of negative information termed as CMNet. A collaboration module (CM) is designed to establish the correlation between the target positive and negative information in a learnable manner via three blocks: target positive, target negative, and interactive block. Experimental results demonstrate our CMNet achieves superior performance than recent methods.
\end{abstract}

\begin{figure*}[htp]
  \centering
  \includegraphics[width=0.8\linewidth]{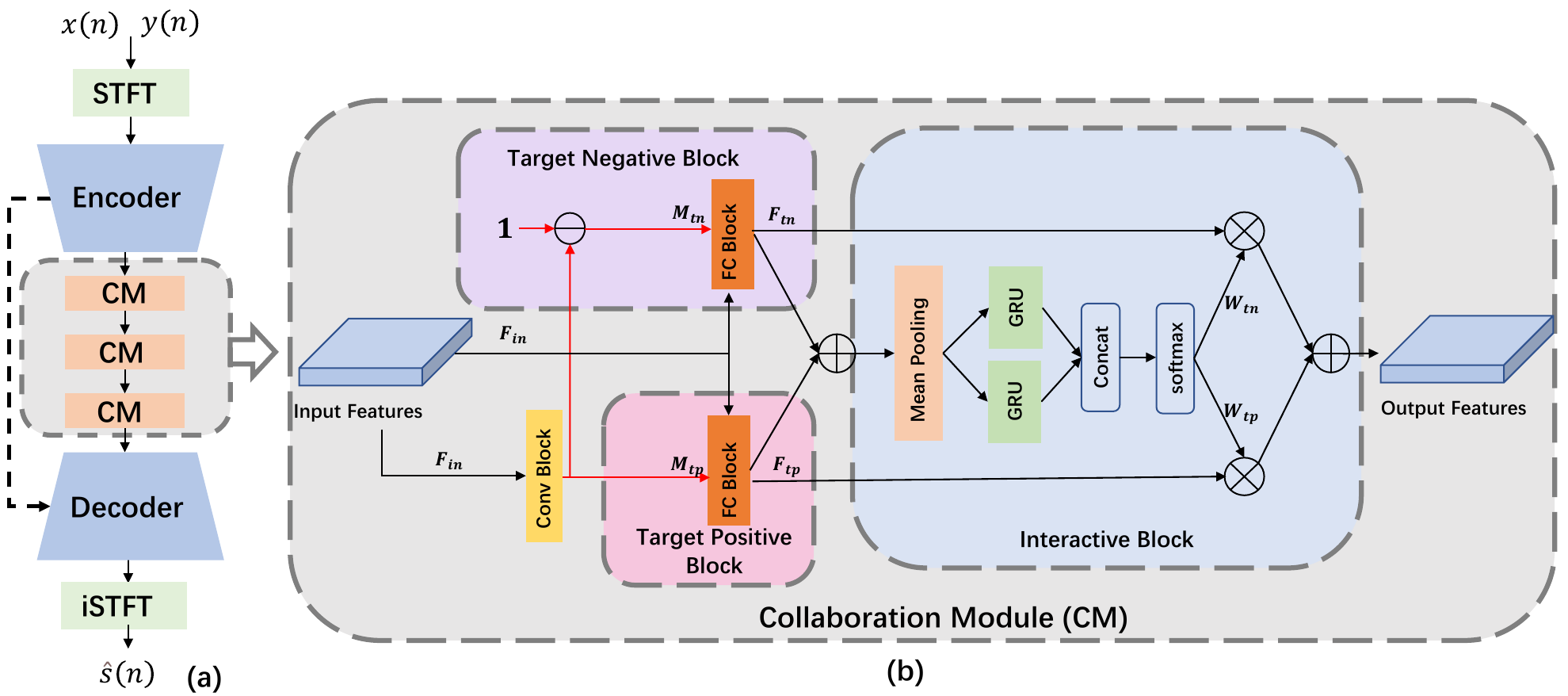}
  \caption{(a) The proposed CMNet. (b) The proposed collaborative module. “FC” denotes feature catcher block. }
  \label{fig:overview}
  \vspace{-0.6cm}
\end{figure*}

\section{Introduction}
Acoustic echo arises in a full-duplex voice communication system when a near-end microphone picks up audio signals from a near-end loudspeaker and sends it back to far-end participants such that the far-end users receive a modified version of their voice. Acoustic echo cancellation (AEC) aims to suppress the echo from the microphone signal while leaving the target near-end signal least distorted.

The traditional AEC applies adaptive filters~\cite{nlms, kalman} to identify the echo path between the loudspeaker and the microphone at the near-end point. The main assumption that underpins conventional AECs is that the echo path is linear \cite{comminiello2010functional}. However, the power amplifiers and loudspeakers, incredibly cheap and small, can be the sources of nonlinearity. When nonlinearities occur, conventional AECs performance decreases the overall achievable quality. Although several nonlinear models \cite{Volterra, Hammerstein, flaf} have been utilized, these traditional methods for convergence may take a lot of work in complex acoustic environments.

Deep learning (DL) has recently gained much attention for its capacity to model complicated nonlinear relationships and has been successfully applied to AEC. DL-based AEC can be formulated as a supervised speech separation/enhancement problem \cite{ss_aec, real-crn-aec, wave-u-net, dtln}, which separates near-end speech from the microphone signal and has shown better performance than their classical counterparts \cite{aec-challenge-2021}. However, DL-based methods often suffer from near-end speech distortion and interference signals (noise and echo) residual, which suggests some indistinguishable patterns between target speech and interference signals are hard to distinguish.

To alleviate this challenge, some approaches introduce target negative information and capture its correlation with target positive information to make the patterns between target and interference signals more discriminative. However, the definition of target positive and negative information is diverse and can be concretely grouped into two categories. The methods of the first category \cite{fazelmultitask, fcrn-2, lideep} aim to predict near-end speech by adding the characteristics of echo signals that are used as target negative information, in which a separate algorithm is set to estimate the echo signal before the target signal prediction. These approaches generally consider the echo signal by incorporating it into loss functions or by directly predicting it as prior knowledge. However, it is difficult to accurately estimate echo signals when faced with strong nonlinear distortion or excessively high noise levels \cite{comminiello2014nonlinear}, which may introduce interfering information for AEC.  

The methods of the second category \cite{dual-path} extract features of near-end speech as target positive information and features of interference signals as negative information by utilizing a two-parallel branch network to model target and interference signals separately. Subsequently, cross-connections between two branches are employed to use the information learned from the other branch to improve the target signal modeling. As a result, the target positive and negative information have interacted in a latent representation space, and the interaction makes the simultaneous modeling of two signals feasible and effective. Although the two-branch approach ensures the preservation of both target speech and interference signal information during the modeling process, cross-connections pass information without any selection resulting in some redundant information. Additionally, the two-branch structure requires many parameters, resulting in high computational costs.

In our study, features whose weights are high and regarded as helpful information to predict the target signal by the neural network are considered as target positive information. In contrast, features whose weight is low and regarded as interference information by the network are considered as target negative information. The aforementioned analysis motivates us to design a new approach with fewer parameters to explore the correlation between target positive and negative information to make the patterns between near-end speech and interference signals more discriminative. We propose a collaboration module (CM) that includes a target positive block, a target negative block, and an interactive block. The target positive/negative block captures the target positive/negative features at local-level from the global-level features. The interactive block integrates the captured features in a self-adaptive and learnable way to establish a correlation between target positive and negative features. Only a single branch structure is required to build the network. In summary, we propose a novel AEC model, CMNet, which inserts CM into the encoder-decoder architecture to achieve superior performance.

The rest of this paper is organized as follows: In section 2, the problem of the AEC is briefly define. Then the model architecture is presented in Section 3. Section 4 is the dataset and experimental settings. Section 5 demonstrates the results and analysis, and a conclusion is shown in Section 6.

\section{Problem formulation}

The microphone signal $y(n)$ is a mixture of echo $d(n)$, near-end speech $s(n)$, and background noise $v(n)$:
\begin{equation}
    y(n) = d(n) + s(n) + v(n) 
\end{equation}
where $n$ is sample index, $d(n)$ is obtained by a linear or nonlinear transform of the far-end signal $x(n)$. Provided that $x(n)$ and $y(n)$ are known, the task of AEC is to estimate near-end signal $\hat{s}(n)$. A time delay compensation module\cite{gcc-phat} based on the generalized cross-correlation phase transform method is used to align the microphone and far-end signal. Our overall model can be formulized as:
\begin{equation}
\vspace{-1.5mm}
\hat{M}=f_{\varphi}\left(X_{r}, Y_{r}, X_{i}, Y_{i}\right)
\vspace{-1.5mm}
\end{equation}
where $f$ and $\varphi$ denotes CMNet and its network parameters, $X$ and $Y$ denote $x(n)$ and $y(n)$ after short-time Fourier transform (STFT) respectively, $r$ and $i$ represent real and imaginary parts of complex spectrogram, $\hat{M}$ is the estimated complex ratio mask (CRM)\cite{crm} optimized by signal approximation, which can be defined as:
\begin{equation}
\mathrm{CRM}=\frac{Y_{r} S_{r}+Y_{i} S_{i}}{Y_{r}^{2}+Y_{i}^{2}}+j \frac{Y_{r} S_{i}-Y_{i} S_{r}}{Y_{r}^{2}+Y_{i}^{2}}
\end{equation}
The final predicted mask of the network $\hat{M}=M_r+j M_i
$ can be defined in polar coordinates $$
\left\{\begin{array}{l}
M_{\text {mag }}=\sqrt{M_r^2+M_i^2} \\
M_{\text {phase }}=\arctan 2\left(M_i, M_r\right)
\end{array}\right.
$$
and the estimated clean speech $\hat{s}$ can be expressed:
\begin{equation}
\hat{s}=Y_{\mathrm{mag}} \cdot M_{\mathrm{mag}} \cdot e^{Y_{\mathrm{phase}}+M_{\text {phase }}}
\end{equation}

\section{Proposed algorithm}
In our design, the proposed CMNet model consists of an encoder-decoder network to learn and reconstruct target signal and a collaboration module network to extract and combine target positive and negative features. The entire CMNet architecture is shown in Fig.\ref{fig:overview}(a). 

\subsection{Encoder and Decoder}
The encoder contains three 2-D convolutional layers with a kernel size of (3, 5). The stride is (1, 1) for the first layer and (1, 2) for the following two. The channel numbers are [16, 32, 64]. The decoder consists of three gated blocks followed by one 2-D convolutional layer\cite{SNnet}. The first layer in the gated block is a deconvolutional layer, followed by a 2-D convolutional block learning multiplicative mask for the corresponding feature from the encoder. The purpose of this mask is to suppress undesired parts of the feature. Afterward, the masked encoder feature is concatenated with the deconvolutional feature and inputted into another 2-D convolutional layer to produce the residual representation. After three gated blocks, the final convolutional layer learns CRM. The kernel size for all 2-D deconvolutional layers is (3,5). The stride is (1,2) for the first two gated blocks and (1,1) for the last. The channel numbers are 32, 16, and 2, respectively. All the 2-D convolutional layers in the decoder have a kernel size of (1,1), a stride of (1,1), and a channel number the same as that of their deconvolutional layers.  All of the convolutional layers in both the encoder and decoder are preceded by a batch normalization (BN) layer and a parametric ReLU (PReLU) activation function. The padding of the convolution layers is set to keep the causality of the whole system.
\begin{figure}[hpt]
  \centering
  \includegraphics[width=0.8\linewidth]{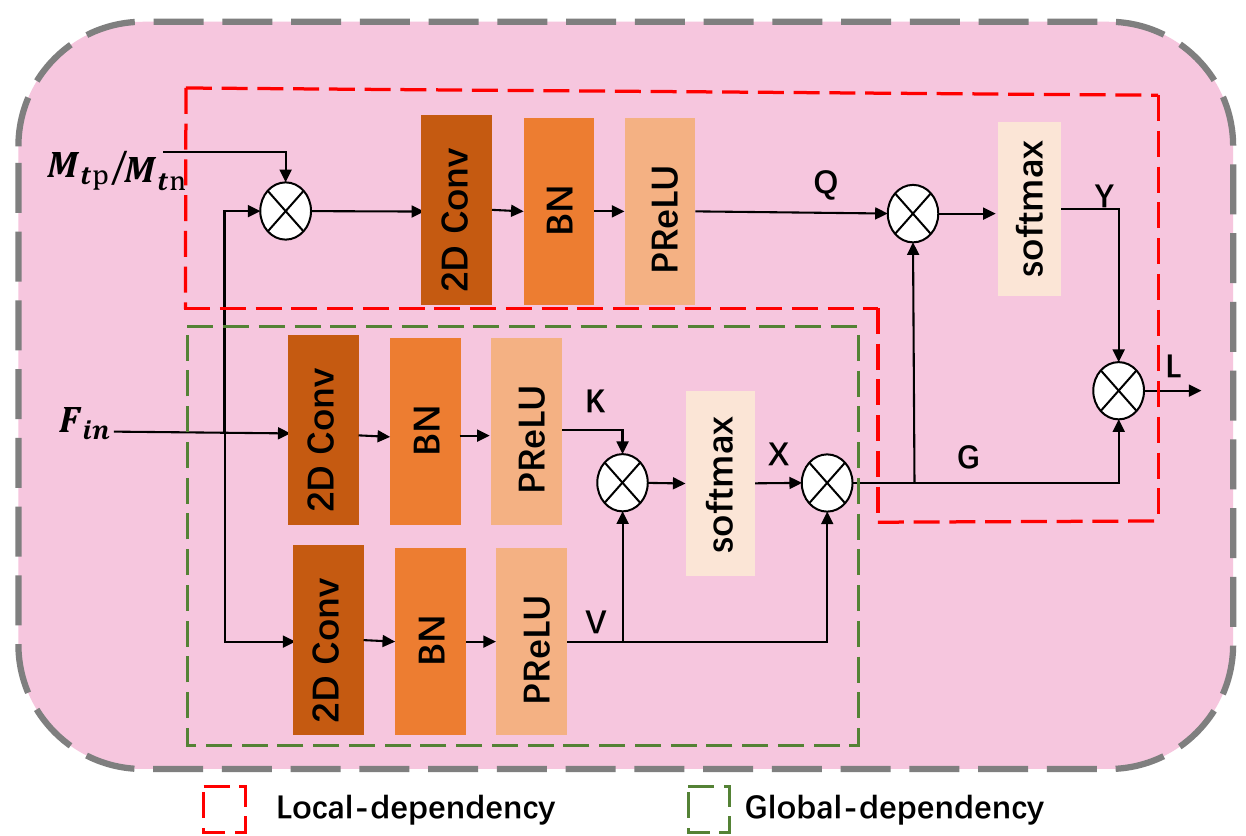}
  \caption{The structure of feature catcher (FC) block}
  \label{fig:gld}
  \vspace{-0.4cm}
\end{figure}
\subsection{Collaboration Module}

Our work presents a crucial idea that target positive and negative information is valuable for AEC to distinguish between target and interference signals. But using these two types of information indiscriminately may result in passive effects. Therefore, we propose a collaborative module (CM) to model and combine target positive and negative features to obtain more discriminative information for AEC. Note that this combination is not a simple series or parallel connection but is learnable and self-adaptive.

The proposed CM consists of three parts: target positive block, target negative block, and interactive block. As demonstrated in Figure\ref{fig:overview}(b), given input features $F_{in}$, a 2D convolutional block is applied to produce the feature map $M_{tp}$, in which the region of target positive features are highlighted, and can be expressed as the following:
\begin{equation}
\mathbf{M}_{t p}=\operatorname{Sigmoid}\left(\operatorname{Conv}\left(\operatorname{ReLU}\left(\operatorname{Conv}\left(\mathbf{F}_{i n}\right)\right)\right)\right)
\end{equation}
After that, we feed $F_{in}$ and $M_{tp}$ into a feature catcher (FC) block, which is designed for capturing local information from global-level features to extract target positive features $F_{tp}$. We utilize one minus $M_{tp}$ to obtain the feature map $M_{tn}$ for target negative features. Similarly, we feed $F_{in}$ and $M_{tn}$ into a FC block to extract target negative features $F_{tn}$. Inspired Li et al.\cite{sknet}, the interactive block is proposed to adaptively integrate $F_{tp}$ and $F_{tn}$. In detail, we first merge $F_{tp}$ and $F_{tn}$ from two branches through an element-wise addition. Next, mean pooling is utilized to generate global features that guide the adaptive selection between target positive and negative features. After that, two gate recurrent units (GRU) are used to generate two weight vectors $w_{tp}$ and $w_{tn}$ for channel-wise selection between two features. $w_{tp}$ and $w_{tn}$ are content-aware, hence learnable and self-adaptive. The interactive block can be expressed as the following steps,  $\mathcal{G}^1$ and $\mathcal{G}^2$ denote two independent GRUs:

\vspace{-0.45cm}
\begin{equation}
\left\{\begin{array}{l}
F_{gf}=MeanPool\left(F_{t n}+F_{t p}\right) \\
w_{tp}, w_{tn}=\operatorname{softmax}\left(\left[\mathcal{G}^1\left(F_{g f}\right), \mathcal{G}^2\left(F_{g f}\right)\right]\right) \\
F_{\text {out }}=W_{t p} \times F_{t p}+W_{t n} \times F_{t n}
\end{array}\right.
\end{equation}

Inspired by Xu et al.\cite{gld}, we propose a FC block to capture speech information in global-level and local-level. As shown in Figure\ref{fig:gld}, we divide the process of computing FC block into two parts: global-dependency and local-dependency. We first capture global dependency by self-attention mechanism. Afterward, we capture the local dependency with the help of global-level features. In global-dependency part, given an input feature $F_{in}$, two 2-D convolutional layers are adopted to generate new feature maps, $\bold{K}$ and $\bold{V}$. The global-dependency part follows the non-local operation\cite{non-local} to compute responses based on relationships between different TF units of the spectrogram.  Consequently, the non-local attention map $\bold{X}$ can be efficiently calculated by dot product:
\begin{equation}
\mathbf{X}=\operatorname{softmax}\left(\mathbf{K} \mathbf{V}^{\top}\right)
\vspace{-4mm}
\end{equation}
The output of global-dependency part $\bold{G}$ is defined as:
\begin{equation}
\mathbf{G}=\mathbf{V} \mathbf{X}^{\top}
\vspace{-3mm}
\end{equation} 
Previously, we extracted feature map $M_{tp}$ at the local level using a convolutional layer with restricted receptive field, which make network pay more attention to target positive features. In local-dependency part, the output of global dependency part $\bold{G}$ and the feature map after the local process $\bold{Q}$ are treated as the inputs.
\begin{equation}
\mathbf{Y}=\operatorname{softmax}\left(\mathbf{Q G}^{\top}\right)
\vspace{-2mm}
\end{equation}
Finally, the output of local-dependency part is computed by the metrics multiplication between $\bold{G}$ and $\bold{Y}$.

\subsection{Training objectives}
We utilize SI-SNR \cite{sisnr} as our loss function:
\begin{equation}
    \boldsymbol{s}_{target} = \frac{<\boldsymbol{\hat{s}},\boldsymbol{s}>\boldsymbol{s}}{||\boldsymbol{s}||^{2}}
\end{equation}
\begin{equation}
    \boldsymbol{e}_{noise} = \boldsymbol{\hat{s}}-\boldsymbol{s}_{target}
\end{equation}
\begin{equation}
    \mathrm{L_{SISNR}} = 10\log_{10}{\frac{||\boldsymbol{s}_{target}||^{2}}{||\boldsymbol{e}_{noise}||^{2}}}
\end{equation}
where $s$ and $\hat{s}$ are the clean and estimated time-domain waveform, respectively. $< ·, · >$denotes the dot product between two vectors and $\Vert . \Vert_2$ is L2 norm. 

\section{Experimental setup}
\begin{table*}[htbp]
\centering
\caption{Ablation experiments. Signal-to-echo ratio(SER) is randomly picked up from $[-15,15]$dB. DT: doubletalk, ST: single-talk, NE: near-end, FE: far-end, TPB: target positive block, TNB: target negative block, IB: interactive block, SNR: signal-to-noise ratio, "-": without additional noise.}
\begin{tabular}{ccccccccccc}
\hline
              & \textbf{}    & \textbf{}    & \textbf{}   & \multicolumn{4}{c}{\textbf{DT}}                                 & \multicolumn{2}{c}{\textbf{ST\_NE}} & \textbf{ST\_FE} \\ \hline
\textbf{}     & \multicolumn{3}{c}{\textbf{SNR(in dB)}}   & \multicolumn{2}{c}{\textbf{-}} & \multicolumn{2}{c}{\textbf{5}} & \multicolumn{2}{c}{\textbf{5}}      & \textbf{-}      \\ \cline{2-11} 
\textbf{CASE} & \textbf{TPB} & \textbf{TNB} & \textbf{IB} & \textbf{PESQ}  & \textbf{STOI} & \textbf{PESE}  & \textbf{STOI} & \textbf{PESQ}    & \textbf{STOI}    & \textbf{ERLE}   \\ \hline
\textbf{1}    & \checkmark            & \checkmark            & \checkmark           & 1.96           & 0.88          & 1.80           & 0.86          & 2.06             & 0.91             & 34              \\
\textbf{2}    & \checkmark            & ×            & ×           & 1.94           & 0.88          & 1.69           & 0.84          & 2.01             & 0.89             & 32              \\
\textbf{3}    & ×            & \checkmark            & ×           & 1.90           & 0.88          & 1.65           & 0.84          & 1.94             & 0.89             & 34              \\
\textbf{4}    & \checkmark            & \checkmark            & ×           & 1.90           & 0.88          & 1.68           & 0.85          & 2.00             & 0.90             & 21              \\
\textbf{5}    & ×            & ×            & ×           & 1.87           & 0.87          & 1.65           & 0.82          & 2.03             & 0.84             & 23              \\ \hline
\end{tabular}
\label{tab:ablation}
\vspace{-4mm}
\end{table*}

\subsection{Dataset}
Our model is trained with 9500 synthetic files from the database provided by Microsoft for the ICASSP 2022 AEC Challenge\cite{2022AEC} that is an open speech corpus.
Besides, we also have performed data expansion.

For data expansion, we first prepare four types of signals: near-end speech, background noise, far-end signal and corresponding echo signal. For near-end speech $s(n)$, there are 10,000 near-end speech utterances in the 2022 official synthetic dataset and we select the first 500 utterances as the test set which is unseen in training. The rest 9,500 utterances, together with 10,000 utterances from 2021 ICASSP AEC-challenge synthetic dataset are used for training.
For background noise $v(n)$, we randomly select 5000 pieces of noise audio from the DNS\cite{dns} dataset for training and 1000 pieces of noise audio for testing. 
For far-end signal $x(n)$ and echo signal $d(n)$, the first 500 sentences of the 2022 official synthetic dataset are used as the test set and the rest 9,500 utterances for training. In addition, we also use the real far-end single-talk utterances provided by the 2021 and 2022 AEC challenge, which covers a variety of recording devices and signal time delay. And then we combine these four signals together to get microphone signal.

\subsection{Implementation details}
All audio signals are resampled to 16kHz. The chunk size of our training data is set to 10s. The proposed model uses STFT to extract the spectrum from each utterance.
A Hamming window with 512 bins and overlap interval of 256 bins is used. Our model is trained with the Adam optimizer with an initial learning rate of 1e-3. For CM, The filter sizes and strides of convolutional layer in the Conv Block of CM are (3, 7) and (1, 1), and in the FC of CM, they are (1, 1) and (1, 1) in time and frequency dimension. We use masks for self-attention and set the GRU to unidirectional in FC to avoid involving future information. The number of hidden units in GRU is 1. The whole parameters of CMNet are 2.5M.

\subsection{Evaluation metrics}

The following three metrics are used to evaluate our model and state-of-the-art competitors. All metrics are better if higher.
\begin{itemize}

\item PESQ: Perceptual evaluation of speech quality (from $-$0.5 to 4.5) \cite{pesq}.
\item STOI: Short-time objective intelligibility measure (from 0 to 1) \cite{stoi}.
\item ERLE: Echo return loss enhancement for far-end single-talk periods \cite{erle}, which is defined as: 
\begin{equation}
\mathrm{ERLE}=10 \log _{10}\left[\sum_{n} y^{2}(n) / \sum_{n} \hat{s}^{2}(n)\right]
\end{equation}
\end{itemize}
PESQ and STOI are used for double-talk and near-end single-talk scenarios and ERLE used for far-end single-talk.

\section{Results and analysis}

We demonstrate the ablation study in Table\ref{tab:ablation} to investigate the effect of different components in CMNet. Note that \textbf{Case 1} denotes the basic CMNet with the default setting. In the ablation study, we compare our method with several baseline models: \textbf{Case 2}, we only keep target positive block of CM. \textbf{Case 3}, we only keep target negative block of CM. \textbf{Case 4}, we remove interactive block in CM. \textbf{Case 5}, we remove whole CM and add self-attention layers to keep the number of parameters roughly constant. A detailed analysis of the ablation study is presented below.
\begin{table}[htbp]

\caption{Echo cancellation performance. SNR is randomly picked up from 5dB or without additional noise in DT and is set to 5dB in NE\_ST. The DT scenario randomly picks SER from [-15,15]dB.}
\resizebox{\linewidth}{!}{
\begin{tabular}{ccccccc}
\hline
                    \textbf{}           & \textbf{}            & \multicolumn{2}{c}{\textbf{DT}} & \multicolumn{2}{c}{\textbf{ST\_NE}} & \textbf{ST\_FE} \\ \cline{2-7} 
\textbf{Method}     & \textbf{\#Params(M)} & \textbf{PESQ}  & \textbf{STOI}  & \textbf{PESQ}    & \textbf{STOI}    & \textbf{ERLE}   \\ 
 \hline
\textbf{CRN}        & 4                    & 1.53                              & 0.83                              & 1.71             & 0.88             & 23              \\
\textbf{F-T-LSTM}   &1.2     &1.67    &0.85  &1.82 &0.89 &23
%\multicolumn{1}{c}{}              & \multicolumn{1}{c}{}              &0.85                  &1.82                  &0.89
\\
\textbf{MTFAA}      & 2.1                  & 1.79                              & \textbf{0.88}                              & 2.00             & \textbf{0.91}             & 25              \\ \hline
\textbf{two-branch} & 4.8                  & 1.74                              & 0.87                              & 1.97             & 0.90             & 28              \\
\textbf{CMNet}      & 2.5                  & \textbf{1.88}                              & \textbf{0.88}                              & \textbf{2.06}             & \textbf{0.91}             & \textbf{34}              \\ \hline
\end{tabular}}
\label{tab:res}
\vspace{-3mm}
\end{table}

\begin{figure}[htp]
  \centering
  \includegraphics[width=\linewidth]{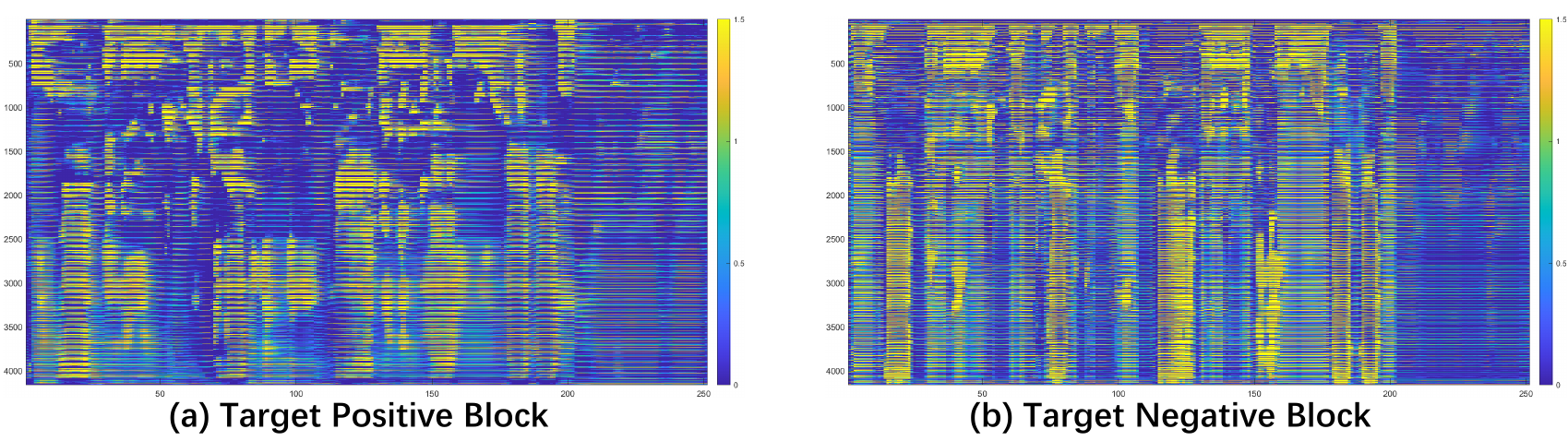}
  \caption{Visualization of attention feature map from different FC blocks. (a) target positive block. (b) target negative block.}
  \label{fig:fea_map}
  
\end{figure}

To verify the effect of target negative information, we make the comparison between Case 1 and Case 2. 0.02 STOI gains and 0.11 PESQ gains by Case 1 over Case 2 at a signal-to-noise ratio of 5 dB show the superiority of the use of target negative, especially in a noisy environment.
In Case 3, we only apply target negative block. The performance comparison between Case 3 and Case 5 indicates the necessity of involving target negative features for the AEC model.
Interestingly, comparing Case 2 with Case 3, we observe that CMNet with only target positive blocks performs slightly better than with only target negative blocks. Figure\ref{fig:fea_map} shows the attention feature maps of target positive and target negative blocks. From (a) and (b), we observe that target positive and target negative blocks behave differently in CM. This is reasonable as the two blocks model different features and their focus differs. And in some indistinguishable patterns between speech and interference signals, the weights in the feature map may be about 0.5, so the feature maps are similar.
In Case 4, we only remove interactive block. The performance comparison between Case 1 and Case 4 suggests the importance of feature fusion in a learnable and adaptive manner.
 
 We compare our CMNet with three other methods, CRN\cite{real-crn-aec}, F-T-LSTM\cite{f-t-lstm}, and MTFAA\cite{mtfaa} trained with our dateset. CRN is a  causal convolutional recurrent network for complex spectral mapping. F-T-LSTM adopts an AEC approach using a complex neural network to better model the important phase information and frequency-time-LSTMs, which scan both frequency and time axis, for better temporal modeling. MTFAA is a system that presents a novel backbone for speech dense-prediction called a multi-scale temporal frequency convolutional network with axial self-attention. Table\ref{tab:res} shows the comparison results in terms of STOI and PESQ. We highlight the best score under each condition in boldface. Besides, based on our network, a two-branch network is constructed by mimicking the structure in ~\cite{dual-path}, which further validates that our model can achieve better results with fewer parameters.

\section{Conclusions}

In this paper, we observe that both the target positive information and the target negative information, such as interference signals and features, are valuable in guiding the AEC model training procedure. Therefore, we propose a novel AEC method termed CMNet, encoder-decoder architecture with a collaboration module (CM) inserted. The CM is used to establish the correlation between the target positive and negative information at the global and local levels in a learnable and self-adaptive manner. Experimental results demonstrate that the guidance of both positive and negative information can improve AEC performance.

\bibliographystyle{IEEEtran}
\bibliography{mybib}

\end{document}